\documentclass[12pt,a4paper]{report}

\usepackage{cvl-titlepage}

\usepackage[pdfborder=0]{hyperref}	% links without border
\usepackage{tabularx}
\usepackage{xcolor}
\usepackage{textcomp}
\usepackage{amsmath}
\usepackage{amsfonts}
\usepackage{mathptmx}
\usepackage{graphicx}

\newcommand{\N}{\mathbb{N}}

\title{Motion Similarity Modeling \\ A State of the Art Report}
\author{Anna Sebernegg, Peter K{\'a}n, Hannes Kaufmann}
\newcommand{\trnumber}{001}	% report number
\abstract{
The analysis of human motion opens up a wide range of possibilities, such as realistic training simulations or authentic motions in robotics or animation.
One of the problems underlying motion analysis is the meaningful comparison of actions based on similarity measures. 
Since the motion analysis is application-dependent, it is essential to find the appropriate motion similarity method for the particular use case.
This state of the art report provides an overview of human motion analysis and different similarity modeling methods, while mainly focusing on approaches that work with 3D motion data.
The survey summarises various similarity aspects and features of motion and describes approaches to measuring the similarity between two actions.
}
\begin{document}

\maketitlepage
\pagenumbering{roman}
\tableofcontents
\clearpage
\pagenumbering{arabic}

%----------------------------------------------
% Introduction
%----------------------------------------------
\chapter{Introduction}
\label{ch:introduction}
Advances in motion capture technologies to digitize human motion have created a foundation for motion analysis.
One main underlying problem of motion analysis is the meaningful comparison of motions by using similarity measures \cite{roder2006similarity}.
%However, no generalized similarity model has yet been found to describe the similarity between human motions suitable for every use-case.
Similarity measures used for the detection and analysis of human motion depend on the context and goal of the intended application \cite{roder2006similarity}. Consequently, it is essential to consider which method and similarity model is suitable.

Related work in the area of motion similarity proposed different similarity models along with algorithms for applications in various fields.
%For example, the detection and analysis of human motion, by using different similarity models, is applied in various fields.
In the medical field, for instance, it can be utilized to track and analyze the process of rehabilitation treatments more precisely and further to determine their success \cite{fern2012biomechanical,zhao2014rule,chang2012towards}.
In Surveillance, the problem of human identification can be addressed by motion recognition. 
Wang et al. have adopted a spatial-temporal silhouette analysis for gait recognition, which can identify people based on their way of walking \cite{wang2003silhouette}.
The analysis of human motion also has its benefits in the field of sports.
It can provide athletes with more detailed feedback and gives insight into comparisons of their movements with other athletes or with themselves. 
In Brodie et al., for example, two ski runs are compared through fusion motion capture and force vector analysis \cite{brodie2008fusion}.
However, similarity modeling may also be used for training purposes, as described by Chan et al. \cite{chan2007immersive}.
Motion similarity models are also increasingly embedded in virtual environments to enable training or rehabilitation exercises at home in a more immersive setting, as described by Chang et al. and Chua et al. \cite{chang2012towards,chua2003training}. 
Similarity models are also applied in robotics, for example, for kinematic mapping or to examine the similarity between humanoid motion and real human motion \cite{zhao2004kinematics}.
Motion indexing and retrieval, motion synthesis, and motion blending are applications that are broadly applied in the entertainment industry, for instance, to create realistic human motions \cite{kruger2008multi}.
In order to perform human motion analysis in such diverse fields, different requirements have to be met -- such as real-time tracking and calculations for virtual reality applications or large amount of data acquisition for sports
%high data acquisition for sports  
-- and various approaches have already been developed \cite{reyes2016human}.
This state of the art report aims to give an overview of several approaches and provides a summary of the advantages and disadvantages of some methods.

Under Chapter \ref{ch:motion}, this paper provides essential information on the human motion as well as on MoCap data.
Chapter \ref{ch:similarityaspects} describes similarity aspects that define when two motions are accounted as similar.
In Chapter \ref{ch:feature}, possible human motion features for similarity modeling are categorized, and brief descriptions are given.
Chapter \ref{ch:comparison} covers proposed methods and algorithms for the comparison of human motion, and the final Chapter \ref{ch:feature} presents a brief conclusion and summarization of the topic.

%----------------------------------------------
% Related Work
%----------------------------------------------
\chapter{Motion}
\label{ch:motion}
Val{\v{c}}{\'\i}k defines motion as a set of trajectories or as a sequence of poses \cite{valvcik2016similarity}.
Natural human motion depends on the three spatial dimensions as well as on time and is influenced by internal and external factors \cite{roder2006similarity}.
The motion of the human skeleton can be described by the following three types of rigid motion as described by Reyes \cite{reyes2016human}:

\begin{itemize}
  \item Linear
  \item Angular
  \item General
\end{itemize}

The linear motion describes a translation in a particular direction over time, while angular motion describes a rotation around a single axis.
The general motion is the combination of both linear and angular motion.

\section{Motion Capture Data}
\label{sec:motion-capturing-data}

There are several different technologies for recording full-body motion in order to perform human motion analysis, including inertial and optical sensors \cite{reyes2016human,roder2006similarity}. 
Primarily appearance-based and model-based approaches are used for the capturing of motion data, as described by Val{\v{c}}{\'\i}k et al. \cite{valvcik2016similarity}.
Appearance-based methods work directly with recorded sequences, like video sequences from a single video camera, where it is possible to extract, for example, silhouettes as data representations \cite{valvcik2016similarity}.
Appearance-based methods usually utilize heuristic assumptions to establish feature correspondence between successive frames \cite{aggarwal1999human}.
Model-based methods, on the other hand, apply predefined models such as skeleton models or volumetric models, which makes it easier to establish feature correspondence and body structure recovery as the tracked data only needs to be matched to the model \cite{aggarwal1999human}.
Because of the complexity of human motion, due to the vast number of degrees of freedom and the limited number of recordable points, the predefined models representing the human body are often simplified, typically by using a skeleton model with restricted sets of joints connected by rigid bones \cite{roder2006similarity}.
The capturing of motion results in motion capture data (MoCap data).
MoCap data formats are not consistently standardized, and thus, various formats, like the proprietary BVH or CSM files, exist \cite{valvcik2016similarity}.

\section{3D Skeleton Representation of MoCap Data}
\label{sec:skeleton}
\begin{figure}[htb]
	\centering
	\includegraphics[width = 0.65\textwidth]{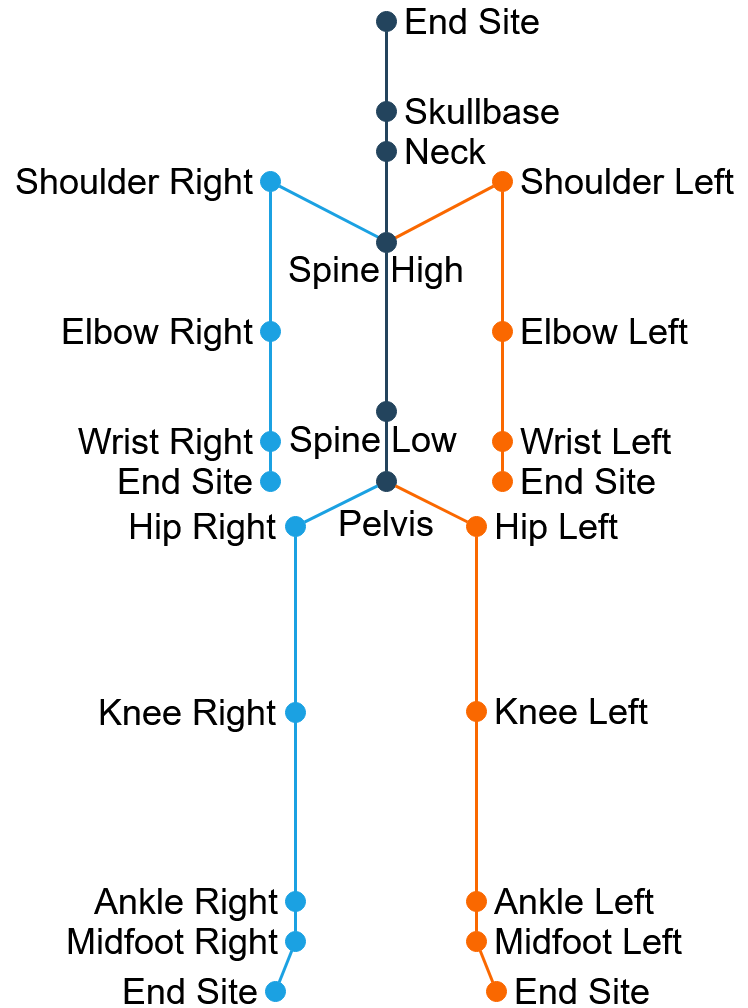}
	\caption[Chain model]{3D Skeletal kinematic chain model. The joints are represented as dots and labeled. The lines connecting the joints represent rigid bones.}
	\label{fig:skeleton}
\end{figure}

%model based
Motion data can be fitted into a skeletal model by transforming the data into a joint chain \cite{wang2016survey}.
As mentioned above, the 3D skeleton representation uses an additional abstraction over the recorded data and provides position, view, and scale invariance due to the known perspective. 
However, calculations based on skeleton data can be computationally intensive if, for example, all joints per frame have to be traversed to derive more features \cite{valvcik2016similarity}. The data representation may also lack smoothness due to motion reconstruction. Therefore, filtering techniques are used to reduce the outlier and noise in motion data \cite{wang2016survey}.
The number of joints and limbs used in a 3D skeleton data set is also dependent upon the application. 
Figure \ref{fig:skeleton} shows one possible 3D skeleton data structure. 

%-------------------------------------------------------
% Methodology
%-------------------------------------------------------
\chapter{Similarity Aspects of Motion}
\label{ch:similarityaspects}

The comparison of similarity between human motions is derived from the context of the intended application due to its ambiguous definition \cite{roder2006similarity}. 
Various factors can play a role, whether a motion is interpreted as similar to another or not.
In some applications, only the rough course of motion may be of interest, while in others, also subtle nuances are viewed as different.
As a result, a central task of motion comparison or analysis is the design of a suitable similarity model; for instance, the selected or designed similarity model should be invariant to arbitrary or tampering aspects \cite{muller2008relational}.
Some of the following similarity aspects may influence the design decisions of a similarity model.

\section{Global Transformations}
\label{sec:global}

In many cases, transformations relative to the origin are not considered a factor of difference in similarity models; accordingly, two motions are considered similar if the only differences are global transformations such as absolute positions in time or space  \cite{roder2006similarity}. 
Global transformations refer not only to translations in the global space but also to rotations about an axis of the global coordinate system or the overall scale/speed of the actor \cite{muller2008relational}.

\section{Content and Style}
\label{sec:contentstyle}

Motion content and style are abstract terms that appear in literature in various forms.
M\"uller et al. refers to \emph{motion style} as the person's individual characteristics/personalized aspects of motion or their performance and emotional expressiveness.
In their paper, they give the example of different walking styles: A walk can be performed in various ways, for example, by tiptoeing, limping, or marching. Different moods can influence the performed motion as well, for example, a walk can seem cheerful or angry.
In contrast, \emph{motion content} is only related to the semantics of motion and thus is close to the raw data \cite{muller2008relational}.
Lee and Elgammal define motion style in the context of gait recognition as the ``time-invariant personalized style of the gait which can be used for identification'', and motion contend as a ``time-dependent factor representing different body poses during the gait cycle'' \cite{lee2004gait}.
For some applications, it may be interesting to separate content and style of the motion, e.g., if the aim is to identify related motions by content independent of motion style. 
This then has to be taken into account in the process of designing the similarity model, for example by utilizing concepts like logically similar motion detection where qualitative features are used to cope with significant
numerical differences in 3D positions or joint angles that can arise through different styles, as described by M\"uller et al. \cite{muller2008relational}.
Lee and Elgammal separated the gait style with a bilinear model to utilize it for gait recognition, and Davis and Gao presented an approach for modeling and recognizing different action styles \cite{lee2004gait,davis2003expressive}.
M\"uller et al. give a further, more detailed overview of how the  concepts of motion content and style are treated in the literature \cite{muller2008relational}.

\section{Logical and Numerical Similarity}
\label{sec:logical}

Logical and numerical similarity are similar concepts to motion content and style.
Kovar et al. define \emph{logical similarity} of motions as ``variations of the same action or sequence of actions'' \cite{kovar2004automated}.
Therefore, logically similar motions share the same action pattern \cite{chen2012partial}.
According to M\"uller et al., the variation of logically similar actions can be influenced by the spatial and the temporal domain. For example, two walking sequences can be logically similar, even when they contain significant spatial and temporal differences \cite{muller2008relational}.
Logical similarity focuses on the content of actions while masking out factors of the individual motion style \cite{roder2006similarity}. 
As an example, different variations of a walking action, resulting from the individual styles of the performers, can still be perceived as logically similar in an application since all of them can be classified as locomotion.

Motions are \emph{numerically similar} if the corresponding numerical values, such as skeleton poses, are approximately the same \cite{kovar2004automated}.
Algorithms, which are developed for numerically similar motion detection, are usually based on numerical/quantitative features that are semantically more close to the raw MoCap data than qualitative descriptions \cite{muller2008relational}.

It is important to note that logical and numerical similarity does not imply each other. 
Kover et al., for example, states that action sequences, which are referred to as logically similar, perhaps be numerically dissimilar and vice versa \cite{kovar2004automated}.
In their publication ``Automated extraction and parameterization of motions in large data sets'', there are also examples given to highlight this concept.

\section{Partial Similarity}
\label{sec:partial}

Partial similarity is the case when certain body parts are moving similarly, while other parts of the body move in different ways \cite{roder2006similarity}. Sometimes partial similarities are more important than the overall similarity of a motion. Integrating extracted features from irrelevant body parts in the similarity measure can then affect the results negatively \cite{chen2012partial}.
M\"uller et al. use a set of Boolean geometry features to express the relations between body parts \cite{muller2005efficient}.
Chen et al. describe a partial similarity motion retrieval based on geometric features \cite{chen2012partial}. As the number of relative geometry features for human motion is vast, they utilized Adaboost for the selection of effective features \cite{chen2012partial}.
Jang et al. synthesize new human body motions by combining different partial motions. They analyze the similarity of partial motions to choose the most natural-looking combinations \cite{jang2008enriching}.
Rule-based approaches, as described by Zhao et al., could perhaps also be used to calculate partial similarity \cite{zhao2014rule}.

%-------------------------------------------------------
% Results
%-------------------------------------------------------
\chapter{Human Motion Features}
\label{ch:feature}

Human motion features, as discussed by Val{\v{c}}{\'\i}k et al., describe different characteristics of human motion and are an abstraction of MoCap data to enable further processing \cite{valvcik2016similarity}.
MoCap data contains information about the position, orientation, and movement of a person in 3D space along with noise.
Depending on the problem definition and the similarity model, some of this information may be irrelevant or even counterproductive for further calculations. 
For example, in cases where semantic/logical similarity is required regardless of the actual global position or orientation of a person.
Therefore, human motion features are derived, which focus only on specific information to leave out unsuitable or misleading data.
The selection and combination of these features then depend on the requirements of the given problem \cite{valvcik2016similarity}. 
It should also be noted that each movement that needs to be distinguished within an application requires an explicit representation by the selected criteria \cite{kamel2019efficient}.
Because distinguishing features may differ for individual actions, accurate action classification and similarity calculations can require case related consideration of various features \cite{moencks2019adaptive}.
Dynamic feature extraction instead of a predefined feature space produced by static feature selection can be of interest for application where various motion types have to be recognized or compared \cite{moencks2019adaptive}.
In static feature selection methods, the subset of features for a feature vector is predefined either manually or by machine learning algorithms and is later not altered. Classifying an action as ``walking'' or ``sitting'' can, for example, be realized by selecting the velocity of the center of mass of the person acting as a feature. However, if multiple different actions, such as sitting activities and dynamic activities such as walking, have to be classified, a simple feature selection may not be sufficient. 
In the given example, the velocity of hip motion in sitting activities would be of minor interest. In contrast, the classification of dynamic activities could potentially benefit from the integration of the hip velocity in the feature vector \cite{moencks2019adaptive}. In static feature selection, each selected feature adds to the dimensionality of the feature space. In contrast, dynamic feature selection adds feature sequentially as needed, i.e., it selects features on the fly according to each new action to be classified \cite{he2013dynamic,cruz2018dynamic}.

Val{\v{c}}{\'\i}k et al. categorizes human motion features into the following four groups:
\nameref{sec:anthropometric}, \nameref{sec:pose}, \nameref{sec:transition} and \nameref{sec:action} \cite{valvcik2016similarity}.
The following subsections describe these categories while mainly focusing on model-based features that concentrate on specific joints.

\section{Anthropometric Features}
\label{sec:anthropometric}

Anthropometric features are quantitative measurements that describe body dimensions of the recorded person, for instance, body height and width, lengths of particular bones such as the arm-length, as well as joint rotation limits and, therefore, are not corresponding to motion.
Some anthropometric information related to the human bone structure can be extracted from 3D skeleton MoCap-Data.
Anthropometric features have minimal use in appearance-based approaches, as opposed to skeleton models, where several features can be derived \cite{valvcik2016similarity}.
Subject-specific features may be helpful in surveillance tasks, such as the recognition and identification of humans, but can be misleading in applications where only the similarity factor between two movements is required. In this case, it is necessary to normalize the 3D skeleton data as addressed by \cite{vox2018preprocessing}.
As an example, the mass of skeleton segments and their center of mass is used by Kr{\"u}ger et al. together with translation features for comparing motions \cite{kruger2008multi}.

\section{Pose Features}
\label{sec:pose}

Pose features describe characteristics of single poses and are extracted from each static posture or single frame independently \cite{agahian2019efficient}. 
Pose features are neither influenced by the speed of action nor by its surrounding frames \cite{valvcik2016similarity}. 
Their time invariance can make pose similarity calculations (e.g., comparing individual key poses) less complicated than the analysis of the overall action, which often involves the analysis of entire pose sequences or the additional use of extra features.
Pose features are used for action recognition tasks as described by Agahian et al. \cite{agahian2019efficient}, as well as for motion-comparison tasks as outlined by Monash et al., where a user has to try to mimic a recorded-gesture displayed on a screen \cite{monash2012motion}.

\subsection{Joint Angle Rotations}
\label{sec:jointanglerotations}

Joint angle rotations are measured on each joint in a 3D skeleton model, and it is important to note that their definition depends on the coordinate system used \cite{valvcik2016similarity}.
Typically local or absolute rotations are used.
Local rotations describe the rotation of an object relative to its parent.
Absolute rotations, on the other hand, describe the rotation of an object relative to the global coordinate system or the coordinate system of the skeletal root.
One can transform local rotations to absolute rotations by hierarchically traversing the skeleton from the root and chaining the corresponding rotations.
In other words, the local coordinate systems are aligned with the global coordinate system in case of the absolute description and rotated relative to the parent in case of the local variant \cite{valvcik2016similarity}.
Rotations can also have different mathematical representations, such as Euler angles, quaternions, rotation matrices, and spherical coordinate systems, where each description has its own distinct properties \cite{huynh2009metrics}.
Furthermore, not all skeleton models are represented by joint rotations. Sometimes joint positions are used. Forward and inverse kinematics can be used in such cases to convert between the two different systems \cite{valvcik2016similarity}.
One major drawback when comparing motions, based only on relative joint rotations, is that rotations in some body parts have a more significant overall effect on a pose or action than others. For example, a small rotation in the shoulder can lead to much larger changes as in the wrist \cite{roder2006similarity}.

\subsection{Distance-Based Pose Features}
\label{sec:distancebased}

Distance-based pose features are based on the joint positions and measure either the distance between two arbitrary joints or between a joint and a defined edge or plane.
The planes for the joint to plane distances can be defined both relative to the subject or absolute.
Anthropometric features influence distances between two joints. Therefore, normalized 3D skeletons usually are used in these cases \cite{valvcik2016similarity}.
An example of joint to joint distance-based features as well as for joint to plane distance measurement -- in this particular case, the global floor is used as an absolute plane -- is given by Ijjina et al., where the features are applied for action recognition \cite{ijjina2014human}.
Two examples for joint to relative plane distances are given in M\"uller et al. and M\"uller and R\"oder \cite{muller2005efficient,muller2006motion}.

\subsection{Relational Features}
\label{sec:relationalfeatures}

\begin{figure}[htb]
	\centering
	\includegraphics[width = 1\textwidth]{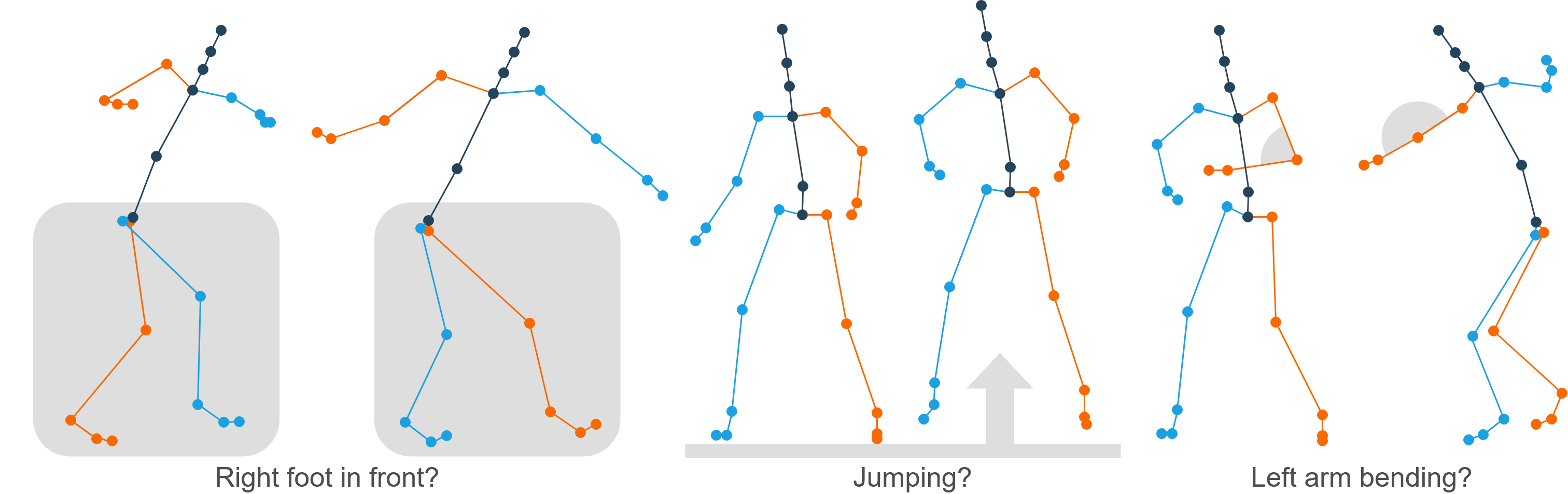}
	\caption[Geometric Features]{Examples of relational features as Boolean expressions, as described by R{\"o}der et al.
  \protect\cite{roder2006similarity}.}
	\label{fig:geometricfeature}
\end{figure}

M\"uller et al. propose relational features as geometric relations between specified body parts or points included in 3D skeleton data \cite{muller2005efficient}.
They provide a semantic representation of motion and are invariant ``to various kinds of spatial deformations of poses'' \cite{roder2006similarity}.
Previously discussed features, such as joint position, are numerical, and thus a quantitative description of motion.
Numerical features can be sensitive towards pose deformations. For this reason, they are not always suitable for logical similarity detection  \cite{thanh2013automatic}. 
Relational features are, on the other hand, a qualitative description that defines an individual or common sequential characteristics of logically similar motions and have the -- for logical similarity favorable --  property of being invariant to local deformations \cite{muller2008relational}.
In M\"uller et al., a set of Boolean geometric features is used \cite{muller2005efficient}. Figure %\ref{fig:geometricfeature} shows some of the Boolean expressions given as an example in their paper, such as `both hands are touching' or `the left hand is in front of the right hand' \cite{muller2005efficient}.
\ref{fig:geometricfeature} shows some possible Boolean expressions, such as `Is the right foot in front of the left foot?' or `Is the left arm bending?'.
In the paper ``Efficient motion search in large motion capture'' written by Yi, more generalized variations of relational features are described, where only features of bones that count as dominant in a specific motion are considered \cite{yi2006efficient}. 

% TODO anna figure

\subsection{Silhouette-Based Pose Features}
\label{sec:silhouette}

The extracted silhouettes from appearance-based models can directly be used as pose features, either as the silhouettes region or its contour \cite{valvcik2016similarity}.
Phillips et al. proposed a baseline algorithm for gait analysis that utilized silhouette as a feature \cite{phillips2002baseline}.
The employment of spatio-temporal silhouette analysis, described by Wang et al., is another example of a gait analysis \cite{wang2003silhouette}. 
Another feature representation by sequences of human
silhouettes is discussed by Huang et al. \cite{huang2010human}.

\section{Transition Features}
\label{sec:transition}

Transition features, as defined by Val{\v{c}}{\'\i}k et al., describe characteristics of transformations/displacements between two or more sequential poses, such as the displacement of joints.
Just like pose features, they are bound to one pose. However, precedent and subsequent poses in a sequence affect the computations as well.
A typical transition feature is the instantaneous velocity \cite{valvcik2016similarity}.

For transition features, it must be considered that joint motion can not only be triggered by the joint itself but also by parented joints, as stated by Kamel et al. 
For example, our fingers can move separately from the rest of our bodies, but they can also move due to the movement of our wrist, elbow, shoulder, or the whole body. When walking, each joint in our body moves along, even if some parts of the body are held steady.
Moreover, most transition features are not explicit. For example, joint position trajectories do not hold information about the direction of movement \cite{kamel2019efficient}. 

Just like joint rotations, joint motion can be distinguished by the applied coordinate system. 
Kamel et al. describe a motion based on the descriptive coordinate system as \emph{local}- or \emph{global}-motion.
They describe the local motion of a joint as the displacement relative to a parental joint while global motion describes the displacement according to a fixed joint (root) or the global coordinate system.
The benefit by using a local coordinate system is that it provides information about the influence of the parent joints, and therefore it is possible to distinguish if a joint moved by itself or was (partly) moved by one of its parents \cite{kamel2019efficient}.

\subsection{Instantaneous Velocity}
\label{sec:ivelocity}

Instantaneous velocity describes the change of joint position or joint rotation between two sample points (i.e., movement of a joint between two frames). Both the absolute and relative velocity apply as transition features, though only the relative velocity describes the relation between velocities of the joints.
Kamel et al. present a motion quantification and subsequent similarity evaluation using both translation and angular velocities from 3D skeletal data with the distinction between local and global motion \cite{kamel2019efficient}.
In Kr{\"u}ger et al., the angular momentum of the body segments, together with the acceleration of the center of mass, is used for comparing motions \cite{kruger2008multi}.
Both the velocity and the acceleration are also used in Moencks et al. as features for action recognition on a multi-modal dataset  \cite{moencks2019adaptive}.
Thanh et al. only utilized the absolute velocity to extract the most active frames and body parts \cite{thanh2013automatic}. 
The velocity vector can also be a description of direction \cite{valvcik2016similarity}.
Velocity usually needs to be derived from MoCap data.
The applied discrete derivation then amplifies quick transition changes that can be caused by measuring errors or noise, which often results in jittering of the skeleton joints.
If no motion smoothing algorithms are applied on the MoCap data, velocity features can lead to significant distortion of the motion description and possibly result in subsequent errors \cite{valvcik2016similarity}.

\subsection{Instantaneous Acceleration}
\label{sec:iacceleration}

Another transition feature is the instantaneous joint acceleration.
The acceleration is derived from velocity and therefore has similar characteristics as velocity features. 
The quality of this feature has, for example, a similar dependency on noise. 
In Kr{\"u}ger et al., the acceleration of the center of mass is used to detect so-called non-contact phases, since the acceleration then corresponds to the acceleration due to gravity \cite{kruger2008multi}.
As mentioned above, acceleration is also a feature utilized in Moencks et al. \cite{moencks2019adaptive}.

\subsection{Kinetic Properties}
\label{sec:kinetic}

The mentioned transition features can be classified as kinematic properties. Kinetic properties, like forces as used in combination with joint angles for measuring motion similarity by Yang et al. \cite{yang2008human}, can be utilized as well.

\section{Action Features}
\label{sec:action}

Action features describe the characteristics of a complete semantic or logical action.
Therefore, the motion sequence has to be analyzed to extract actions beforehand.
This extraction is either done by user input or automatically based on other features \cite{valvcik2016similarity}. 
For instance, an action could be delimited by two key poses.
Examples for action features are the duration of action, the total displacement of joints, periodicity patterns such as walk cycles and rhythms of motion as well es the average velocity and average acceleration \cite{valvcik2016similarity}.

The main difference to the instantaneous velocity described in Section \ref{sec:transition} is that the average velocity not only represents a short momentum between two frames but averages it over a whole sequence or total action.
The average velocity, therefore, is the total displacement divided by the total time of the action.

In other publications, statistical descriptions, such as the mean, median, modus, standard deviation, or minimum or maximum values of, e.g., acquired joint angle velocity, are utilized as action features \cite{valvcik2016similarity}. An example is given by Ball et al., where the k-means algorithm is used for gait recognition \cite{ball2012unsupervised}.
Other widely adopted features are trajectories in two- or three-dimensional space \cite{reyes2016human}. Trajectories describe how a coordinate or value evolves and can be displayed graphically as a curve diagram.
The typical joint trajectories, for example, represent the path a joint follows through space as a function of time. Joint-angle trajectories, as utilized by Zhao et al. as well as by Tanawongsuwan and Bobick, represent the change of joint-angles over time \cite{tanawongsuwan2001gait,zhao2004humanoid}.

%-------------------------------------------------------
% Comparison
%-------------------------------------------------------

\chapter{Human Motion Comparison}
\label{ch:comparison}

Motion similarity analysis is application orientated, therefore different approaches where developed over the time to meet the particular requirements \cite{tang2008emulating}. 
The similarity between poses and actions can be measured by using distance metrics or learning methods \cite{gavrila1999visual}. Additional tasks that build on the measured similarity, such as action recognition, depend heavily on the accuracy of the distance metric or learning process \cite{porikli2004trajectory}.
This section gives an overview of a selection of approaches and focuses mainly on the different choices of feature vectors, preprocessing of the data, and similarity measurements.

\section{Data Preprocessing - Dimensionality Reduction}
\label{sec:reduction}

The analysis of human motion and the calculation of similarity are complicated by the high dimensionality and complexity of the motion data \cite{martinez2017human,witte2009motion}. 
Since it is not easy to work with motion data in its raw form, different approaches are utilized for simplifying the data by dimensionality reduction and filtering \cite{forbes2005efficient}.
Two popular methods, Principal Component Analysis and Self Organizing Map are described below.

\subsection{Principal Component Analysis}
\label{sec:pca}

Principal Component Analysis (\acs{pca}) is a linear method to simplifying a multivariate dataset by reducing data dimensionality \cite{lippi2011can}.
The \acs{pca} and other projection methods attempt to find the best approximated subspace for the data set to which the data can be projected (typically in terms of variance) \cite{witte2009motion}.
The execution of a \acs{pca} on a given dataset leads to a vector space of equal dimensions where each axis in space represents a principal component vector. Points in space are then weighted combination of principal components \cite{forbes2005efficient}.
Data reduction is obtained by using only a subset of its principal components \cite{forbes2005efficient,lippi2011can}.
Several authors such as R{\"o}der and Tido \cite{roder2006similarity}, Agahian et al. \cite{agahian2019efficient} or Witte et al. \cite{witte2009motion} apply \acs{pca} before the comparison of motion data to achieve dimensionality reduction.
K. Forbes and E. Fiume used a weighted \acs{pca}-based pose representation for pose-to-pose distance calculations to find similar motions in a database \cite{forbes2005efficient}.

\subsection{Self Organizing Map}
\label{sec:som}

A Self Organizing Map (\acs{som}) is a neural network that is trained by unsupervised learning and produces a low-dimensional and discrete representation of the input space. Therefore \acs{som}s are used for dimension reduction \cite{huang2010human,sakamoto2004motion}.
Sakamoto et al. map large data sets onto a two-dimensional discrete space by the help of \acs{som}s \cite{sakamoto2004motion}.
Huang and Wu also utilize this technology for their human action recognition approach to reduce data dimensionality as well as to cluster feature data \cite{huang2010human}.
They use sequences of human silhouettes as the primary feature and extract key poses through the trained \acs{som}. As a similarity measure, they utilize the Euclidean distance and later the longest common subsequence (LCS) method for comparing trajectories \cite{huang2010human}.

\section{Local Similarity Measures}
\label{sec:local}

Single frames can hold a lot of information if they contain an expressive pose, such as a kick or hit of a tennis player. 
Local similarity measures compare such individual poses and, therefore, do not take temporal aspect into account \cite{roder2006similarity}. 
They can be used for comparing key-poses or are incorporated into more complex, time-dependent similarity models.
Popular pose distance functions are for example the \textbf{Manhattan distance (L1)}, \textbf{Euclidean distance (L2)} and the \textbf{Cosine distance} \cite{valvcik2016similarity}.
Chan et al., for example, utilize the cosine similarity as a local similarity measure for each pair of joint angles in a frame \cite{chan2007immersive}.
They proposed an immersive performance training tool where trainees have to imitate the simulated trainer.
Their movement was compared and analyzed by posture matching, to be able to give the trainees feedback \cite{chan2007immersive}.  
In addition to the drawback of needing to define thresholds as error tolerance, this method also introduces the problem of not being invariant to global rotations if absolute and not relative angles are used. 
Other distance metrics for rotations can be utilized if joint rotations are represented as unit quaternions, such as the \textbf{total weighted quaternion distance} or \textbf{geodesic distance} \cite{roder2006similarity}.
%Six different distance functions for rotations are also analyzed by Huynh \cite{huynh2009metrics}. 
The following distance functions for rotations are analyzed in more detail by Huynh \cite{huynh2009metrics}:

\subsection{Euclidean Distance between Euler Angles}
\label{sec:euclidean}

The Euclidean distance applies the Pythagorean theorem to $p$ dimensions. 
For 2D, this distance measure can be interpreted as the length of the chord between two Euler angles on the unit circle.
Euler angles are not unique, i.e., several Euler angles can represent the same rotation. However, their Euclidean distance may result in a non-zero value, which does not reflect the actual distance between them. 
The Euclidean distance between Euler angles can also lead to large distance values between nearby rotations, while two distant rotations may lead to smaller values.
Therefore, the paper does not recommend this metric for calculating the difference between the two rotations \cite{huynh2009metrics}.
Switonski et al. stated that distance functions based on quaternions allow a more efficient assessment of rotation similarities, in comparison to Euler angles \cite{switonski2012dynamic}.

\subsection{Norm of the Difference of Quaternions}
\label{sec:norm}

Unit quaternions give a more flexible representation of rotations. Other than Euler angles, their values do not dependent on the order of rotations about the three principal axes and do not suffer the gimbal lock problem. They are well suited for interpolations as well \cite{switonski2012dynamic}. 
The norm of the difference of quaternions is one of many distance functions defined in quaternion space. 
This metric $\Phi$ defines the distance between two rotations as the Euclidean distance between two unit quaternions $q1$ and $q2$ as given in Equation \ref{equ:distance2} \cite{huynh2009metrics}.
As unit quaternions $q$ and $-q$ denote the same rotation, the minimum operator is required \cite{huggins2014comparing}.

\begin{equation}
   \Phi(q_1,q_2) = min(||q_1 - q_2||, ||q_1 + q_2||).
\label{equ:distance2}
\end{equation}

\subsection{Inner Product of Unit Quaternions}
\label{sec:inner}

A similar metric for calculating the distance between unit quaternions is given by the inner product as denoted in Equation \ref{equ:distance3} \cite{huynh2009metrics}.
This metric is, for example, used by Wunsch et al. for 3D object pose estimation \cite{wunsch1997real}.

\begin{equation}
   \Phi(q_1,q_2) = min(arccos(dot(q_1, q_2)), \pi - arccos(dot(q_1, q_2))).
\label{equ:distance3}
\end{equation}

The metric can alternatively be replaced by the following computationally more efficient functions:

\begin{equation}
   \Phi(q_1,q_2) = arccos(|dot(q_1, q_2)|).
\label{equ:distance4}
\end{equation}

\begin{equation}
   \Phi(q_1,q_2) = 1 - |dot(q_1, q_2)|.
\label{equ:distance5}
\end{equation}

\subsection{Deviation from the Identity Matrix}
\label{sec:deviation}

Distance functions can be based of matrix representations as well. 
One drawback is however, that they usually are more computationally expensive. The following metric given as Equation \ref{equ:distance6} tries to find the amount of rotation required to align a rotation matrix $R_1$ with $R_2$.
Among the distance metrics described by Huynh, this metric is the most computationally expensive metric. However, the computation work can be significantly reduced if unit quaternions instead of matrices are used \cite{huynh2009metrics}.
\begin{equation}
   \Phi(R_1,R_2) = ||I - R_1R_2^T||_F.
\label{equ:distance6}
\end{equation}

\subsection{Geodesic on the Unit Sphere}
\label{sec:geodesic}

This metric, as denoted in Equation \ref{equ:distance7}, gives a geodesic on the unit sphere, which can be interpreted as the shortest curve between two rotations lying on the surface of the unit sphere.
As shown by Huynh, this metric has a linear relationship with Equation \ref{equ:distance4}, and therefore can be calculated more simply based on unit quaternions.
\begin{equation}
   \Phi(R_1,R_2) = ||log(R_1R_2^T)||.
\label{equ:distance7}
\end{equation}

Arikan and Forsyth give a more complex example for the usage of a local similarity measure, by utilizing multiple features like joint positions, velocities, and accelerations \cite{arikan2002interactive}.
A lot of similarity metrics, such as the one proposed by Arikan and Forsyth, utilize different joint weights to control the relevance of different joints \cite{yang2005motion}.
One major drawback of such metrics is the need to define the optimal attribute weights for the given problem definitions and the dependence of the quality of results on selected weights \cite{yang2005motion}.

Kovar et al. employ a model based approach working on \textbf{point clouds} to measure the similarity between two frames \cite{kovar2008motion}. 
The point cloud distance metric is defined as the minimum difference between two point sets (a squared form of L2 distance) \cite{valvcik2016similarity}. 
Each point cloud is the composition of smaller point clouds that represent the pose at each frame in a defined window of neighboring frames \cite{kovar2008motion}. 
Drawbacks of the point cloud approach described by Kovar et al. are the coordinate-invariance and the efficiency \cite{yang2005motion}.

%---------------------------------------------------------------------------------------------------

\section{Global Similarity Measures}
\label{sec:globameasure}

\subsection{Temporal Alignments}
\label{sec:temporal}

In image-based human pose comparison, the similarity can be assessed through conventional approaches such as measuring the distance between joint positions or rotations \cite{coskun2018human}. These approaches are well suited, for example, to determine the similarity between detected key-poses. However, when human poses have to be compared over a time span, assessing the similarity between two poses or motions becomes a non-trivial problem \cite{coskun2018human}.
Humans tend to perform movements slightly differently each time, which means that certain poses appear at different frames in sequences that represent the same human motion \cite{forbes2005efficient,coskun2018human}.
Such temporal changes are subtle for a human observer. However, they can lead to an enormous numerical difference when performing computer-aided motion analysis \cite{forbes2005efficient}.
Nevertheless, two actions must often be identified as similar motions, even if they are spatio-temporal variations of each other \cite{muller2008relational}.
Therefore, different types of algorithms, such as dynamic time warping, are often used in motion data processing to obtain a temporal match of the used motions.
The result of a time warp depends not only on the algorithm itself but also on the utilized distance function and the compared features \cite{kruger2008multi}. 
Dimensionality reduction allows for computational advantages. Similarity models based on machine learning, for example, can achieve a shorter training time by reducing the amount of data necessary to train \cite{mandery2016dimensionality}.

Pose-to-pose comparison is also utilized in global similarity measures, for example, in combination with time alignment methods.
Forbes and Fiume give an example of a similarity model based on both local and global distance measures \cite{forbes2005efficient}. 
They present a weighted-\acs{pca} based pose representation in combination with Euclidean distance metric \cite{forbes2005efficient}.
\acs{pca} trajectories are then compared by using the Dynamic Time Warping (\acs{dtw}) distance \cite{roder2006similarity}.
A similar algorithm for constructing a content-based human motion retrieval system is proposed by Chiu et al. \cite{chiu2004content}.
They compute the similarity between the query example and each candidate clip through \acs{dtw} on \acs{som}-based
trajectory clusters \cite{chiu2004content,roder2006similarity}.

\subsubsection{Dynamic Time Warping}
\label{sec:dtw}

\begin{figure}[htb]
	\centering
	\includegraphics[width = 0.6\textwidth]{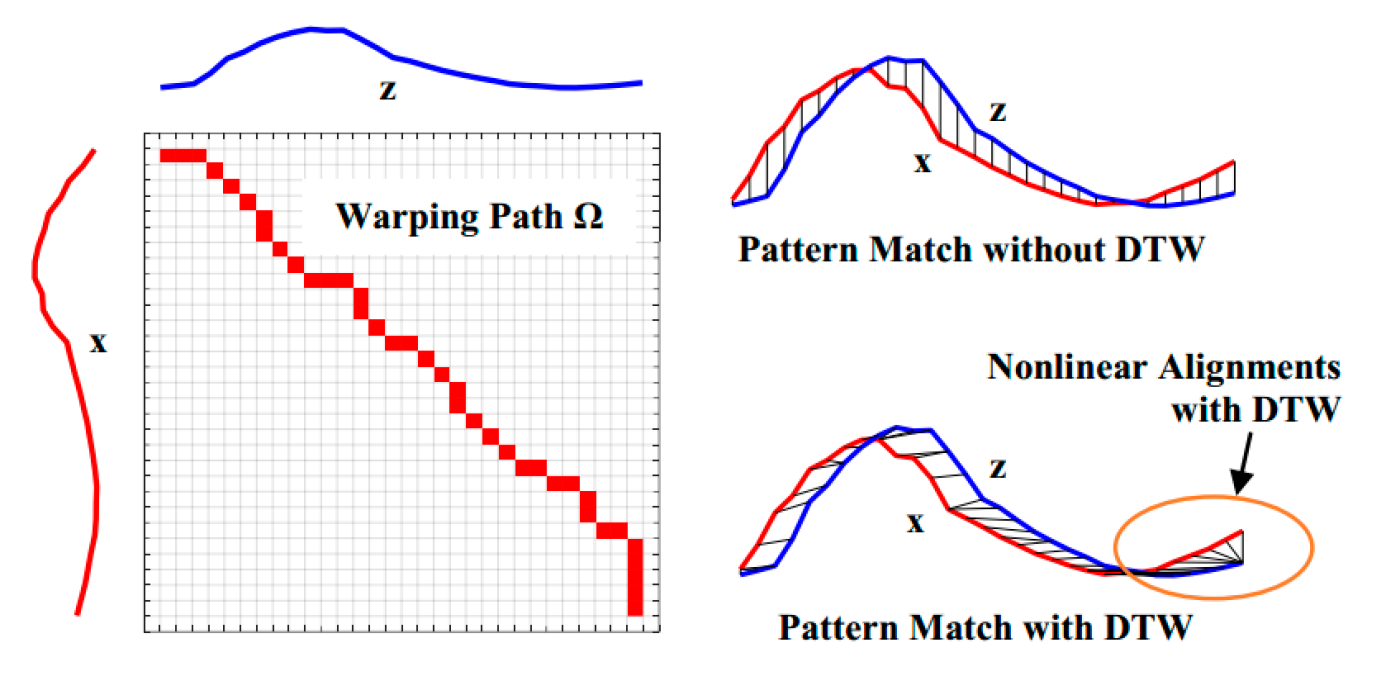}
	\caption[\acs{dtw}]{Dynamic time warping. Illustration by Yang et al. \cite{yang2019imu}.}
	\label{fig:dtw}
\end{figure}

%% computational intensiv
Dynamic Time Warping (\acs{dtw}) determines the optimal alignment between two given temporal sequences and can be used for measuring similarity between them as it allows the comparison of two time-series sequences with varying lengths and speeds \cite{reyes2016human}.
%%%%
Given two time series $X = (x_1, x_2, ..., x_N), N \in \N$ and $Y = (y_1, y_2, ..., y_M), M \in \N$ with equidistant points in time and sequence sizes of $N$ and $M$.
To align both sequences, \acs{dtw} starts by calculating the local cost matrix $C \in \mathbb{R}^{NxM}: c_{i,j} = ||x_i - y_j||, i \in [1:N], j \in [1:M]$ that represents all pairwise distances between $X$ and $Y$ \cite{senin2008dynamic}.
The algorithm then finds an alignment or warping path from the first cell [1,1] to the last cell [N, M], which runs through the low-cost areas on the local cost matrix. To find the optimal warping path $\Omega$ of the minimum total cost as shown in Figure \ref{fig:dtw}, one would need to test every warping path between $X$ and $Y$, which would be computationally expensive. Therefore, \acs{dtw} utilizes discrete dynamic programming to build an accumulated cost matrix $D$ in a recursive fashion where the warping path $\Omega$ can be found by following the greedy strategy. This optimization leads to a total complexity of O(NM) \cite{senin2008dynamic,folgado2018time,yang2019imu}.
The warping path can then be employed to align the series in time.
%%%%%
Similarity models using the default \acs{dtw} implementation on high-dimensional data have, however, two main disadvantages: the quadratic complexity and the limitation in capturing the semantic relationship between two sequences, by disregarding the temporal context \cite{coskun2018human}.
\acs{dtw} is still a widely adopted solution for time alignment problems.
M\"uller and R\"oder, for example, employed \acs{dtw} in combination with template matching for action recognition \cite{muller2006motion}.
Kr\"uger et al. used an extended variant called 'Iterative multi-scale dynamic time distortion' (\acs{imdtw}), which, in contrast to the basic \acs{dtw}, does not require a quadratic runtime and memory space \cite{kruger2008multi}.
Reyes refers to FastDTW as a less computationally intensive alternative to \acs{dtw} as it is linear in both time and space complexity \cite{reyes2016human}. Another methods with linear complexity are Uniform Time Warping (\acs{utw}) and its extension named Interpolated Uniform Time Warping (\acs{iutw}) \cite{valvcik2016similarity}.
Other examples for \acs{dtw}-based similarity models are given by \cite{chiu2004content,forbes2005efficient}.

%% TODO FIGURE anna

\subsubsection{Hidden Markov Models}
\label{sec:hmm}

Another method for matching time-varying data and action recognition are Hidden Markov Models (\acs{hmm}), which are statistical models that are based on hidden states that represent an activity \cite{aggarwal2011human}.
In contrast to \acs{dtw}, \acs{hmm}s involve a training and classification stage \cite{gavrila1999visual}.
\acs{hmm}s also disregard the temporal context by assuming that observations are temporally independent \cite{wang2016survey}.
\acs{hmm} is used by Yamato et al. for the matching of human motion \cite{yamato1992recognizing} while
Mandery et al. utilized it for the reduction of feature space for whole-body human
action recognition tasks \cite{mandery2016dimensionality}.
Porikli proposed \acs{hmm}-based distance metrics to determine the similarity between trajectories \cite{porikli2004trajectory}.

%\paragraph{Template Matching}
%"The notion of template matching is simple: compare the input pattern with prestored
%patterns in the database. The computational cost of the template matching
%technique is lower than that of other two techniques. However, it is more sensitive
%to the variance of the movement duration."
%TODO röder
%"The notion of template matching is simple: compare the input pattern with pre-stored patterns in the database." \cite{chiu2004content}

\acs{dtw} and \acs{hmm} are widely adopted solutions for the analysis of time-varying data, but other methods for time alignment are used as well. 
Val{\v{c}}{\'\i}k proposed \textbf{Uniform Scaling} (\acs{us}), \textbf{Scaled and Warp Matching} (combination of \acs{us} and \acs{dtw}) and \textbf{Move Split Merge} \cite{valvcik2016similarity}.
D. M. Gavrila referenced approaches based on \textbf{Neural Networks} (\acs{nn}) and also mentioned the possibility to disregard the time component of human motion data by using representations in different spaces such as a \textbf{phase-space} \cite{gavrila1999visual}.
In ``Content-based retrieval for human motion data'' template matching is discussed as a time-alignment method, where input patterns are compared with pre-stored patterns in a database \cite{chiu2004content}.
M{\"u}ller et al. achieve \textbf{spatio-temporal invariance} in their proposed concept by working with geometric features and adaptive-temporal segmentation \cite{muller2005efficient}.

%---------------------------------------------------------------------------------------------------

\subsection{Similarity Based on Curve Representations}
\label{sec:curve}

%ature of Trajectories
Trajectory-based approaches interpret an activity as a set of space-time trajectories. The similarity is then measured between the trajectories to archive motion analysis and action recognition \cite{aggarwal2011human,agahian2019efficient}. 
Trajectories can be captured through motion capturing techniques with markers or through the space-time movement of 3d skeleton data. The whole set of joint-trajectories then represents the full-body motion \cite{reyes2016human}.
Several approaches use trajectories themselves as motion representation or as further human motion features, as discussed in Section \ref{ch:feature} \cite{aggarwal2011human}.

Trajectory based approaches can achieve a detailed analysis of human motion and in many cases are view-invariant \cite{aggarwal2011human}. However, they dependent on well reconstructed or captured data.
Trajectories can also be used to extract further features \cite{porikli2004trajectory,aggarwal2011human}.
Reyes, for example, proposed a human motion model where trajectories are mapped into chain codes by using orthogonal changes of direction. Chain codes allow data
reduction and the use of string matching algorithms \cite{reyes2016human}.

A simple metric to compare a pair of trajectories is the mean of coordinate distances (Cartesian distance, L-norm, ...), which can be enhanced by further statistical values such as median, variance, minimum, or maximum distance \cite{porikli2004trajectory}. 
As mentioned by Porikli, these trajectory distance metrics have the disadvantage that they depend on mutual coordinate correspondences and therefore are limited to compare only trajectories of equal duration (unless they are normalized) \cite{porikli2004trajectory}. In the paper, it is also noted that normalization destroys the temporal properties of the trajectory. Porikli, therefore, proposed an \acs{hmm}-based distance metric to compare trajectories of different temporal properties \cite{porikli2004trajectory}.
Another action recognition problem solved by comparing trajectory, as mentioned before, is proposed by Huang et al. \cite{huang2010human}.

\subsection{Rule Based Approaches}
\label{sec:rule}

Similarity models are often template-based and focus on gesture or action recognition. Rule-based approaches, on the other hand, are primarily used to analyze the correctness of motion, for example, in the context of rehabilitation exercise monitoring. 
Rule-based similarity models define a ground-truth by a set of rules and compare the observed motion with them instead of comparing it with a previously recorded reference motion.
The usage of predefined rules can simplify the implementation of real-time feedback to the patient \cite{zhao2014rule}.

\subsection{Other Mentionable Methods}
\label{sec:other}

There also exist approaches to measure the similarity of 3D models or single poses by comparing skeleton graphs or other tree-based representations. Brennecke et al., for instance, employ graph similarity for 3D shape matching while Chen et al. \cite{brennecke20043d} proposed a novel skeleton tree representation matching approach \cite{chen2017research}.

Kovar and Gleicher identified numerically similar motions based on a match web, which is a DTW-based index structure \cite{kovar2004automated}.
Parisi et al. use self-organizing Growing When Required networks (GWR) to achieve noise-tolerant action recognition in realtime \cite{parisi2015self}.
Nomm and Buhhalko proposed a system for monitoring and supervising therapeutic exercises based on a neural network. The network is used for motion prediction and will interrupt an exercise if certain predefined thresholds are exceeded \cite{nomm2013monitoring}. 
%-------------------------------------------------------
% Conclusion
%-------------------------------------------------------
\chapter{Conclusion}
\label{ch:conclusion}

This state of the art report provides a brief overview of multiple approaches for human motion analysis and similarity modeling.
It must be noted, however, that only a small insight into the vast number of methods is given. In recent years, many papers on human motion analysis have been released to address problems such as real-time action recognition, motion retrieval, or monitoring the correct execution of an action.
For all these tasks, the similarity between actions or at least individual poses must be measured.
The proposed approaches for similarity modeling are diverse and focus on different aspects. They are based on various distinct definitions of similarity, depending on their area of application, and utilize a different combination of features.
What the models have in common, however, is the high dimensional and spatiotemporal-varying MoCap-data on which they are built.
Multiple approaches, therefore, utilize techniques for dimensionality reduction and time alignments such as the wide adopted Principal Component Analysis or Dynamic Time Warping. 
However, by employing such techniques, it must be ensured that they do not result in any information loss or high runtime.
As a consequence, numerous approaches adopt popular and promising algorithms, such as Dynamic Time Warping, and modify them to their specific purpose. 

Most of the discussed approaches are template-based, where an observed motion is compared with a pre-recorded one. However, rule-based methods or the combination of both could also be of interest for applications in which feedback is provided over the execution of an action.

% acronyms & appendix
%\include{appendix}
\chapter*{List of acronyms}
\label{chap:acros}
\addcontentsline{toc}{chapter}{List of Acronyms}

%\chapter{Acronyms-TEST}
%Über die Befehle \acs{betai} wird dann nur das Symbol bzw. die Abkürzung im Text 
%gedruckt. \acf{betai} gibt zusätzlich noch die Erklärung aus, \acl{betai} fügt 
%lediglich die Beschreibung ein.

% BEISPIEL
% \acro{sift}[SIFT]{Scale Invariant Feature Transform}

\begin{acronym}[LONGACR]
\setlength{\itemsep}{\parsep}

\acro{cc} 		[\textsc{cc}]				{Connected Component}
\acro{cv} 		[\textsc{cv}]				{Computer Vision}
\acro{dog}		[DoG]								{Difference-of-Gaussian}
\acro{dtw}		[Dtw]								{Dynamic Time Warping}
\acro{fast}		[\textsc{Fast}]			{Features from Accelerated Segment Test}
\acro{gloh}		[\textsc{Gloh}]			{Gradient Location-Orientation Histogram}
\acro{hmm} 		[\textsc{Hmm}]			{Hidden Markov Model}
\acro{ica}		[\textsc{Ica}]			{Independent Component Analysis}
\acro{id3}		[ID3]								{Iterative Dichotomiser 3}
\acro{imdtw}	[\textsc{Imdtw}]		{Iterative Multi-scale Dynamic Time Distortion}
\acro{jpg}		[\textsc{Jpeg}]			{Joint Photographic Experts Group}
\acro{knn} 		[\textit{k}-\textsc{nn}] 	{\textit{k}-Nearest Neighbor}
\acro{log}		[LoG]								{Laplacian-of-Gaussians}
\acro{mser}		[\textsc{Mser}] 		{Maximally Stable Extremal Regions}
\acro{nn}			[NN]								{Neural Networks}
\acro{np-hard}[\textsc{np}-hard]	{\textbf{n}on-deterministic \textbf{p}olynomial-time hard}
\acro{ocr}		[\textsc{Ocr}]			{Optical Character Recognition}
\acro{pca}		[\textsc{Pca}]			{Principal Component Analysis}
\acro{pda} 		[\textsc{Pda}] 			{Personal Digital Assistant}
\acro{rbf}		[\textsc{Rbf}]			{Radial Basis Function}
\acro{sift}		[\textsc{Sift}]			{Scale Invariant Feature Transform}
\acro{som}		[\textsc{Som}]			{Self Organizing Map}
\acro{surf}		[\textsc{Surf}]			{Speeded Up Robust Features}
\acro{susan}	[\textsc{Susan}]		{Smallest Univalue Segment Assimilating Nucleus}
\acro{svm}		[\textsc{Svm}]			{Support Vector Machine}
\acro{vc}			[\textsc{Vc}]				{Vapnik-Chervonenkis}
\acro{us}			[\textsc{Us}]				{Uniform Scaling}
\acro{utw}			[\textsc{Utw}]				{Uniform Time Warping}
\acro{iutw}			[\textsc{Iutw}]				{Interpolat Uniform Time Warping}

\end{acronym}

% bibliography
\addcontentsline{toc}{chapter}{Bibliography}
\bibliographystyle{alpha}
\bibliography{cvl_technical_report}

\newcommand{\etalchar}[1]{$^{#1}$}
\begin{thebibliography}{MDSRK19}

\bibitem[AC99]{aggarwal1999human}
Jake~K Aggarwal and Quin Cai.
\newblock Human motion analysis: A review.
\newblock {\em Computer vision and image understanding}, 73(3):428--440, 1999.

\bibitem[AF02]{arikan2002interactive}
Okan Arikan and David~A Forsyth.
\newblock Interactive motion generation from examples.
\newblock {\em ACM Transactions on Graphics (TOG)}, 21(3):483--490, 2002.

\bibitem[ANK19]{agahian2019efficient}
Saeid Agahian, Farhood Negin, and Cemal K{\"o}se.
\newblock An efficient human action recognition framework with pose-based
  spatiotemporal features.
\newblock {\em Engineering Science and Technology, an International Journal},
  2019.

\bibitem[AR11]{aggarwal2011human}
Jake~K Aggarwal and Michael~S Ryoo.
\newblock Human activity analysis: A review.
\newblock {\em ACM Computing Surveys (CSUR)}, 43(3):1--43, 2011.

\bibitem[BI04]{brennecke20043d}
Angela Brennecke and Tobias Isenberg.
\newblock 3d shape matching using skeleton graphs.
\newblock In {\em SimVis}, pages 299--310. Citeseer, 2004.

\bibitem[BRRV12]{ball2012unsupervised}
Adrian Ball, David Rye, Fabio Ramos, and Mari Velonaki.
\newblock Unsupervised clustering of people from'skeleton'data.
\newblock In {\em Proceedings of the seventh annual ACM/IEEE international
  conference on Human-Robot Interaction}, pages 225--226, 2012.

\bibitem[BWP08]{brodie2008fusion}
Matthew Brodie, Alan Walmsley, and Wyatt Page.
\newblock Fusion motion capture: a prototype system using inertial measurement
  units and gps for the biomechanical analysis of ski racing.
\newblock {\em Sports Technology}, 1(1):17--28, 2008.

\bibitem[CCD{\etalchar{+}}03]{chua2003training}
Philo~Tan Chua, Rebecca Crivella, Bo~Daly, Ning Hu, Russ Schaaf, David Ventura,
  Todd Camill, Jessica Hodgins, and Randy Pausch.
\newblock Training for physical tasks in virtual environments: Tai chi.
\newblock In {\em IEEE Virtual Reality, 2003. Proceedings.}, pages 87--94.
  IEEE, 2003.

\bibitem[CCW{\etalchar{+}}04]{chiu2004content}
Chih-Yi Chiu, Shih-Pin Chao, Ming-Yang Wu, Shi-Nine Yang, and Hsin-Chih Lin.
\newblock Content-based retrieval for human motion data.
\newblock {\em Journal of visual communication and image representation},
  15(3):446--466, 2004.

\bibitem[CHL{\etalchar{+}}17]{chen2017research}
Xin Chen, Jingbin Hao, Hao Liu, Zhengtong Han, and Shengping Ye.
\newblock Research on similarity measurements of 3d models based on skeleton
  trees.
\newblock {\em Computers}, 6(2):17, 2017.

\bibitem[CJTC{\etalchar{+}}18]{coskun2018human}
Huseyin Coskun, David Joseph~Tan, Sailesh Conjeti, Nassir Navab, and Federico
  Tombari.
\newblock Human motion analysis with deep metric learning.
\newblock In {\em Proceedings of the European Conference on Computer Vision
  (ECCV)}, pages 667--683, 2018.

\bibitem[CLTK07]{chan2007immersive}
Jacky Chan, Howard Leung, Kai~Tai Tang, and Taku Komura.
\newblock Immersive performance training tools using motion capture technology.
\newblock In {\em Proceedings of the First International Conference on
  Immersive Telecommunications}, page~7. ICST (Institute for Computer Sciences,
  Social-Informatics and~…, 2007.

\bibitem[CLZ{\etalchar{+}}12]{chang2012towards}
Chien-Yen Chang, Belinda Lange, Mi~Zhang, Sebastian Koenig, Phil Requejo, Noom
  Somboon, Alexander~A Sawchuk, and Albert~A Rizzo.
\newblock Towards pervasive physical rehabilitation using microsoft kinect.
\newblock In {\em 2012 6th international conference on pervasive computing
  technologies for healthcare (PervasiveHealth) and workshops}, pages 159--162.
  IEEE, 2012.

\bibitem[CSC18]{cruz2018dynamic}
Rafael~MO Cruz, Robert Sabourin, and George~DC Cavalcanti.
\newblock Dynamic classifier selection: Recent advances and perspectives.
\newblock {\em Information Fusion}, 41:195--216, 2018.

\bibitem[CSLL12]{chen2012partial}
Songle Chen, Zhengxing Sun, Yi~Li, and Qian Li.
\newblock Partial similarity human motion retrieval based on relative geometry
  features.
\newblock In {\em 2012 Fourth International Conference on Digital Home}, pages
  298--303. IEEE, 2012.

\bibitem[DG03]{davis2003expressive}
James~W Davis and Hui Gao.
\newblock An expressive three-mode principal components model of human action
  style.
\newblock {\em Image and Vision Computing}, 21(11):1001--1016, 2003.

\bibitem[FBM{\etalchar{+}}18]{folgado2018time}
Duarte Folgado, Mar{\'\i}lia Barandas, Ricardo Matias, Rodrigo Martins, Miguel
  Carvalho, and Hugo Gamboa.
\newblock Time alignment measurement for time series.
\newblock {\em Pattern Recognition}, 81:268--279, 2018.

\bibitem[FBSL12]{fern2012biomechanical}
Adso Fern'ndez-Baena, Antonio Sus{\'\i}n, and Xavier Lligadas.
\newblock Biomechanical validation of upper-body and lower-body joint movements
  of kinect motion capture data for rehabilitation treatments.
\newblock In {\em 2012 fourth international conference on intelligent
  networking and collaborative systems}, pages 656--661. IEEE, 2012.

\bibitem[FF05]{forbes2005efficient}
Kevin Forbes and Eugene Fiume.
\newblock An efficient search algorithm for motion data using weighted pca.
\newblock In {\em Proceedings of the 2005 ACM SIGGRAPH/Eurographics symposium
  on Computer animation}, pages 67--76, 2005.

\bibitem[Gav99]{gavrila1999visual}
Dariu~M Gavrila.
\newblock The visual analysis of human movement: A survey.
\newblock {\em Computer vision and image understanding}, 73(1):82--98, 1999.

\bibitem[HDIE13]{he2013dynamic}
He~He, Hal Daum{\'e}~III, and Jason Eisner.
\newblock Dynamic feature selection for dependency parsing.
\newblock In {\em Proceedings of the 2013 conference on empirical methods in
  natural language processing}, pages 1455--1464, 2013.

\bibitem[Hug14]{huggins2014comparing}
David~J Huggins.
\newblock Comparing distance metrics for rotation using the k-nearest neighbors
  algorithm for entropy estimation.
\newblock {\em Journal of computational chemistry}, 35(5):377--385, 2014.

\bibitem[Huy09]{huynh2009metrics}
Du~Q Huynh.
\newblock Metrics for 3d rotations: Comparison and analysis.
\newblock {\em Journal of Mathematical Imaging and Vision}, 35(2):155--164,
  2009.

\bibitem[HW10]{huang2010human}
Wei Huang and QM~Jonathan Wu.
\newblock Human action recognition based on self organizing map.
\newblock In {\em 2010 IEEE International Conference on Acoustics, Speech and
  Signal Processing}, pages 2130--2133. IEEE, 2010.

\bibitem[IM14]{ijjina2014human}
Earnest~Paul Ijjina and C~Krishna Mohan.
\newblock Human action recognition based on mocap information using convolution
  neural networks.
\newblock In {\em 2014 13th International Conference on Machine Learning and
  Applications}, pages 159--164. IEEE, 2014.

\bibitem[JLLL08]{jang2008enriching}
Won-Seob Jang, Won-Kyu Lee, In-Kwon Lee, and Jehee Lee.
\newblock Enriching a motion database by analogous combination of partial human
  motions.
\newblock {\em The Visual Computer}, 24(4):271--280, 2008.

\bibitem[KG04]{kovar2004automated}
Lucas Kovar and Michael Gleicher.
\newblock Automated extraction and parameterization of motions in large data
  sets.
\newblock {\em ACM Transactions on Graphics (ToG)}, 23(3):559--568, 2004.

\bibitem[KGP08]{kovar2008motion}
Lucas Kovar, Michael Gleicher, and Fr{\'e}d{\'e}ric Pighin.
\newblock Motion graphs.
\newblock In {\em ACM SIGGRAPH 2008 classes}, pages 1--10. 2008.

\bibitem[KSL{\etalchar{+}}19]{kamel2019efficient}
Aouaidjia Kamel, Bin Sheng, Ping Li, Jinman Kim, and David~Dagan Feng.
\newblock Efficient body motion quantification and similarity evaluation using
  3-d joints skeleton coordinates.
\newblock {\em IEEE Transactions on Systems, Man, and Cybernetics: Systems},
  2019.

\bibitem[KTMW08]{kruger2008multi}
Bj{\"o}rn Kr{\"u}ger, Jochen Tautges, Meinard M{\"u}ller, and Andreas Weber.
\newblock Multi-mode tensor representation of motion data.
\newblock {\em JVRB-Journal of Virtual Reality and Broadcasting}, 5(5), 2008.

\bibitem[LC11]{lippi2011can}
Vittorio Lippi and Giacomo Ceccarelli.
\newblock Can principal component analysis be applied in real time to reduce
  the dimension of human motion signals?
\newblock In {\em BIO Web of Conferences}, volume~1, page 00055. EDP Sciences,
  2011.

\bibitem[LE04]{lee2004gait}
Chan-Su Lee and Ahmed Elgammal.
\newblock Gait style and gait content: bilinear models for gait recognition
  using gait re-sampling.
\newblock In {\em Sixth IEEE International Conference on Automatic Face and
  Gesture Recognition, 2004. Proceedings.}, pages 147--152. IEEE, 2004.

\bibitem[MBR17]{martinez2017human}
Julieta Martinez, Michael~J Black, and Javier Romero.
\newblock On human motion prediction using recurrent neural networks.
\newblock In {\em Proceedings of the IEEE Conference on Computer Vision and
  Pattern Recognition}, pages 2891--2900, 2017.

\bibitem[MDSRK19]{moencks2019adaptive}
Mirco Moencks, Varuna De~Silva, Jamie Roche, and Ahmet Kondoz.
\newblock Adaptive feature processing for robust human activity recognition on
  a novel multi-modal dataset.
\newblock {\em arXiv preprint arXiv:1901.02858}, 2019.

\bibitem[Mon12]{monash2012motion}
H~Ali Monash.
\newblock Motion comparison using microsoft kinect.
\newblock {\em computer science project, MONASH UNIVERSITY}, 2012.

\bibitem[MPBA16]{mandery2016dimensionality}
Christian Mandery, Matthias Plappert, J{\'u}lia Borras, and Tamim Asfour.
\newblock Dimensionality reduction for whole-body human motion recognition.
\newblock In {\em 2016 19th International Conference on Information Fusion
  (FUSION)}, pages 355--362. IEEE, 2016.

\bibitem[MR06]{muller2006motion}
Meinard M{\"u}ller and Tido R{\"o}der.
\newblock Motion templates for automatic classification and retrieval of motion
  capture data.
\newblock In {\em Proceedings of the 2006 ACM SIGGRAPH/Eurographics symposium
  on Computer animation}, pages 137--146, 2006.

\bibitem[MR08]{muller2008relational}
Meinard M{\"u}ller and Tido R{\"o}der.
\newblock A relational approach to content-based analysis of motion capture
  data.
\newblock In {\em Human Motion}, pages 477--506. Springer, 2008.

\bibitem[MRC05]{muller2005efficient}
Meinard M{\"u}ller, Tido R{\"o}der, and Michael Clausen.
\newblock Efficient content-based retrieval of motion capture data.
\newblock In {\em ACM SIGGRAPH 2005 Papers}, pages 677--685. 2005.

\bibitem[NB13]{nomm2013monitoring}
Sven Nomm and Kirill Buhhalko.
\newblock Monitoring of the human motor functions rehabilitation by neural
  networks based system with kinect sensor.
\newblock {\em IFAC Proceedings Volumes}, 46(15):249--253, 2013.

\bibitem[Por04]{porikli2004trajectory}
Fatih Porikli.
\newblock Trajectory distance metric using hidden markov model based
  representation.
\newblock In {\em IEEE European Conference on Computer Vision, PETS Workshop},
  volume~3. Citeseer, 2004.

\bibitem[PSR{\etalchar{+}}02]{phillips2002baseline}
P~Jonathon Phillips, Sudeep Sarkar, Isidro Robledo, Patrick Grother, and Kevin
  Bowyer.
\newblock Baseline results for the challenge problem of humanid using gait
  analysis.
\newblock In {\em Proceedings of Fifth IEEE International Conference on
  Automatic Face Gesture Recognition}, pages 137--142. IEEE, 2002.

\bibitem[PWW15]{parisi2015self}
German~I Parisi, Cornelius Weber, and Stefan Wermter.
\newblock Self-organizing neural integration of pose-motion features for human
  action recognition.
\newblock {\em Frontiers in neurorobotics}, 9:3, 2015.

\bibitem[Rey16]{reyes2016human}
Francisco Javier~Torres Reyes.
\newblock {\em Human motion: analysis of similarity and dissimilarity using
  orthogonal changes of direction on given trajectories}.
\newblock University of Colorado at Colorado Springs, 2016.

\bibitem[R{\"o}d06]{roder2006similarity}
Tido R{\"o}der.
\newblock {\em Similarity, retrieval, and classification of motion capture
  data}.
\newblock PhD thesis, 2006.

\bibitem[Sen08]{senin2008dynamic}
Pavel Senin.
\newblock Dynamic time warping algorithm review.
\newblock {\em Information and Computer Science Department University of Hawaii
  at Manoa Honolulu, USA}, 855(1-23):40, 2008.

\bibitem[SKK04]{sakamoto2004motion}
Yasuhiko Sakamoto, Shigeru Kuriyama, and Toyohisa Kaneko.
\newblock Motion map: image-based retrieval and segmentation of motion data.
\newblock In {\em Proceedings of the 2004 ACM SIGGRAPH/Eurographics symposium
  on Computer animation}, pages 259--266, 2004.

\bibitem[SMJ{\etalchar{+}}12]{switonski2012dynamic}
Adam Switonski, Agnieszka Michalczuk, Henryk Josinski, Andrzej Polanski, and
  Konrad Wojciechowski.
\newblock Dynamic time warping in gait classification of motion capture data.
\newblock In {\em Proceedings of World Academy of Science, Engineering and
  Technology}, number~71, page~53. World Academy of Science, Engineering and
  Technology (WASET), 2012.

\bibitem[TB01]{tanawongsuwan2001gait}
Rawesak Tanawongsuwan and Aaron Bobick.
\newblock Gait recognition from time-normalized joint-angle trajectories in the
  walking plane.
\newblock In {\em Proceedings of the 2001 IEEE Computer Society Conference on
  Computer Vision and Pattern Recognition. CVPR 2001}, volume~2, pages II--II.
  IEEE, 2001.

\bibitem[TCKL13]{thanh2013automatic}
Tran~Thang Thanh, Fan Chen, Kazunori Kotani, and Bac Le.
\newblock Automatic extraction of semantic action features.
\newblock In {\em 2013 International Conference on Signal-Image Technology \&
  Internet-Based Systems}, pages 148--155. IEEE, 2013.

\bibitem[TLKS08]{tang2008emulating}
Jeff~KT Tang, Howard Leung, Taku Komura, and Hubert~PH Shum.
\newblock Emulating human perception of motion similarity.
\newblock {\em Computer Animation and Virtual Worlds}, 19(3-4):211--221, 2008.

\bibitem[Val16]{valvcik2016similarity}
Jakub Val{\v{c}}{\'\i}k.
\newblock Similarity models for human motion data.
\newblock {\em Ph. D. dissertation}, 2016.

\bibitem[VW18]{vox2018preprocessing}
Jan~P Vox and Frank Wallhoff.
\newblock Preprocessing and normalization of 3d-skeleton-data for human motion
  recognition.
\newblock In {\em 2018 IEEE Life Sciences Conference (LSC)}, pages 279--282.
  IEEE, 2018.

\bibitem[Wan16]{wang2016survey}
Qifei Wang.
\newblock A survey of visual analysis of human motion and its applications.
\newblock {\em arXiv preprint arXiv:1608.00700}, 2016.

\bibitem[WSP09]{witte2009motion}
K~Witte, H~Schobesberger, and C~Peham.
\newblock Motion pattern analysis of gait in horseback riding by means of
  principal component analysis.
\newblock {\em Human movement science}, 28(3):394--405, 2009.

\bibitem[WTNH03]{wang2003silhouette}
Liang Wang, Tieniu Tan, Huazhong Ning, and Weiming Hu.
\newblock Silhouette analysis-based gait recognition for human identification.
\newblock {\em IEEE transactions on pattern analysis and machine intelligence},
  25(12):1505--1518, 2003.

\bibitem[WWH97]{wunsch1997real}
Patrick Wunsch, Stefan Winkler, and Gerd Hirzinger.
\newblock Real-time pose estimation of 3d objects from camera images using
  neural networks.
\newblock In {\em Proceedings of International Conference on Robotics and
  Automation}, volume~4, pages 3232--3237. IEEE, 1997.

\bibitem[Y{\etalchar{+}}08]{yang2008human}
Yi-Ting Yang et~al.
\newblock Human recognition based on kinematics and kinetics of gait.
\newblock In {\em The 13th International Conference on Biomedical Engineering
  Suntec Singapore International Convention \& Exhibi, Suntec, Singapore},
  2008.

\bibitem[YCWJ19]{yang2019imu}
Chan-Yun Yang, Pei-Yu Chen, Te-Jen Wen, and Gene~Eu Jan.
\newblock Imu consensus exception detection with dynamic time warping -- a
  comparative approach.
\newblock {\em Sensors}, 19(10):2237, 2019.

\bibitem[YG05]{yang2005motion}
Herb Yang and Tong Guan.
\newblock Motion similarity analysis and evaluation of motion capture data.
\newblock 2005.

\bibitem[Yi06]{yi2006efficient}
LT~Yi.
\newblock Efficient motion search in large motion capture database.
\newblock {\em ISVC}, pages 151--160, 2006.

\bibitem[YOI92]{yamato1992recognizing}
Junji Yamato, Jun Ohya, and Kenichiro Ishii.
\newblock Recognizing human action in time-sequential images using hidden
  markov model.
\newblock In {\em CVPR}, volume~92, pages 379--385, 1992.

\bibitem[ZHD{\etalchar{+}}04]{zhao2004humanoid}
Xiaojun Zhao, Qiang Huang, Peng Du, Dongming Wen, and Kejie Li.
\newblock Humanoid kinematics mapping and similarity evaluation based on human
  motion capture.
\newblock In {\em International Conference on Information Acquisition, 2004.
  Proceedings.}, pages 426--431. IEEE, 2004.

\bibitem[ZHPL04]{zhao2004kinematics}
Xiaojun Zhao, Qiang Huang, Zhaoqin Peng, and Kejie Li.
\newblock Kinematics mapping and similarity evaluation of humanoid motion based
  on human motion capture.
\newblock In {\em 2004 IEEE/RSJ International Conference on Intelligent Robots
  and Systems (IROS)(IEEE Cat. No. 04CH37566)}, volume~1, pages 840--845. IEEE,
  2004.

\bibitem[ZLER14]{zhao2014rule}
Wenbing Zhao, Roanna Lun, Deborah~D Espy, and M~Ann Reinthal.
\newblock Rule based realtime motion assessment for rehabilitation exercises.
\newblock In {\em 2014 IEEE Symposium on Computational Intelligence in
  Healthcare and e-health (CICARE)}, pages 133--140. IEEE, 2014.

\end{thebibliography}

\end{document}